\documentclass[preprint,aps ,nofootinbib]{revtex4}
\usepackage{graphicx}
\pdfoutput=1
\usepackage{epsfig}
\usepackage{amsmath}
\usepackage{amsfonts}
\usepackage{amssymb}
\usepackage{bm}
\usepackage{color}
\usepackage{dcolumn}
\usepackage{hyperref}
\usepackage{multirow}
\usepackage{booktabs}
\usepackage{soul}
\usepackage{ulem}
\usepackage{physics}
\providecommand{\U}[1]{\protect\rule{.1in}{.1in}}

\newcommand{\f}{\begin{equation}}
\newcommand{\ff}{\end{equation}}
\newcommand{\fa}{\begin{eqnarray}}
\newcommand{\ffa}{\end{eqnarray}}

\begin{document}
\title{The generation rate of quantum gravity induced entanglement with multiple massive particles}
\author{Pan Li $^{1,2}$}
\email{lipan@ihep.ac.cn}
\author{Yi Ling $^{1,2}$}
\email{lingy@ihep.ac.cn} 
\author{Zhangping Yu $^{1,2}$}
\email{yuzp@ihep.ac.cn} 
\affiliation{$^1$Institute of High Energy
Physics, Chinese Academy of Sciences, Beijing 100049, China\\ $^2$
School of Physics, University of Chinese Academy of Sciences,
Beijing 100049, China }

\begin{abstract}
We investigate the generation rate of quantum gravity induced entanglement of masses(QGEM) in setup with multiple quantum massive particles, among of which only the gravity interaction due to the Newton potential is taken into account. When the distance between any two adjacent Stern-Gerlach (SG) devices is fixed, we consider all the possible configurations of the setup with the same number of particles. In particular, we systemically analyze the case of particle number n=4 and find that the 
prism setup with a massive particle at the center is the most efficient setup for the entanglement generation. This result can be extended to a system with multiple particles up to seven, where the entanglement efficiency is also enhanced in comparison with the setup with fewer particles. This work provides the strategy to construct the QGEM setup with the best generation rate of entanglement.

\end{abstract}
\maketitle

\section{Introduction}
The quantum theory of gravity as one of the most important problems in modern physics has been investigated for decades \cite{Polchinski:1998rq}\cite{Rovelli:2004tv}\cite{Maldacena:1997re}. Various possible effects of quantum gravity have been explored at the phenomenological level as well. Although a lot of considerations based on the theoretical consistency require that the nature of gravity should be quantum and must exhibit some characteristic behavior in contrast to the other known quantum interactions. Unfortunately, until now no such effects have been observed in experiments yet, due to  the weakness of gravity. Even worse, from the perspective of experiment, we could not safely say that gravity must be quantum. The lack of experimental evidence leads to a lot of debates on whether the nature of gravity is quantum or classical\cite{Rosenfeld:1963hjy}\cite{Penrose:1996cv}\cite{Carlip:2008zf}.  For a recent review we refer to \cite{Carney:2018ofe}.

 Recently, a novel strategy has been proposed in \cite{Bose:2017nin}\cite{Marletto:2017kzi} to test the quantum nature of gravity just at the low energy level. By the virtue of local operations and classical communication(LOCC), they design a setup with two Stern-Gerlach devices in which two massive particles in superposition states of position may be entangled by Newton gravity potential. LOCC claims that classical force can not generate quantum entanglement between two particles. If initially two massive particles without entanglement becomes entangled in the final states after passing through the Stern-Gerlach device, then it means the gravity which serves as the unique medium between them must be quantum, even though the gravity is weak enough to be described well by Newtonian potential.  QGEM experiment has opened a new window for testing the quantum effect of gravity and stimulated a lot of  discussions on low energy effects of quantum gravity, such as testing the superposition states of spacetime from quantum geometry point of view \cite{Christodoulou:2018cmk} and the discreteness of time\cite{Christodoulou:2018xiy}; testing gravity-induced reduction of quantum states\cite{Howl:2018qdl};  exploring the quantum nature of the Newtonian potential \cite{Belenchia:2018szb}-\cite{Danielson:2021egj}; testing  the non-locality of quantum gravity \cite{Buoninfante:2018xiw} as well as detecting the quantum gravity effects by Non-Gaussianity\cite{Howl:2020isj}. Some other relevant work on this theme can be found in \cite{Marshman:2019sne}-\cite{Miki:2020hvg}.  
 
 Currently, in order to make the entanglement more easily to be observed, one central issue on QGEM experiment itself is to improve the setup such that the entanglement between massive particles could be generated with the best efficiency.  To guarantee that the entanglement between different particle states is strong enough to be observed within the time regime that the superposition of states could be remained, theoretically the phase difference of states should become large as soon as possible during the evolution. The specific parameter estimation has been performed in \cite{Bose:2017nin}, where it requires that the mass of the matter wave reaches at least $10^{-14}$ kg, and the interaction duration of two matter waves should keep above 2.5 s. That is to say, under these conditions the generated phase difference would be large enough to verify the existence of the entanglement between two particles. However, reaching such parameter regimes is still a big challenge in lab. Therefore, theoretically one is urged to present novel schemes to improve the QGEM setup so as to loose the constraints on experiment parameters and make the experiment easier to implement. In this route one key issue is to   reduce the time duration for generating observable entanglement. With this success then one may weaken the bound for the mass of the particles which would make the experiment 
more practical. Such efforts have been made in \cite{Nguyen:2019huk}\cite{Schut:2021svd}, where more different setups of experiment are designed to improve the original QGEM setup.

\begin{figure}
\begin{center}
\includegraphics[scale=1]{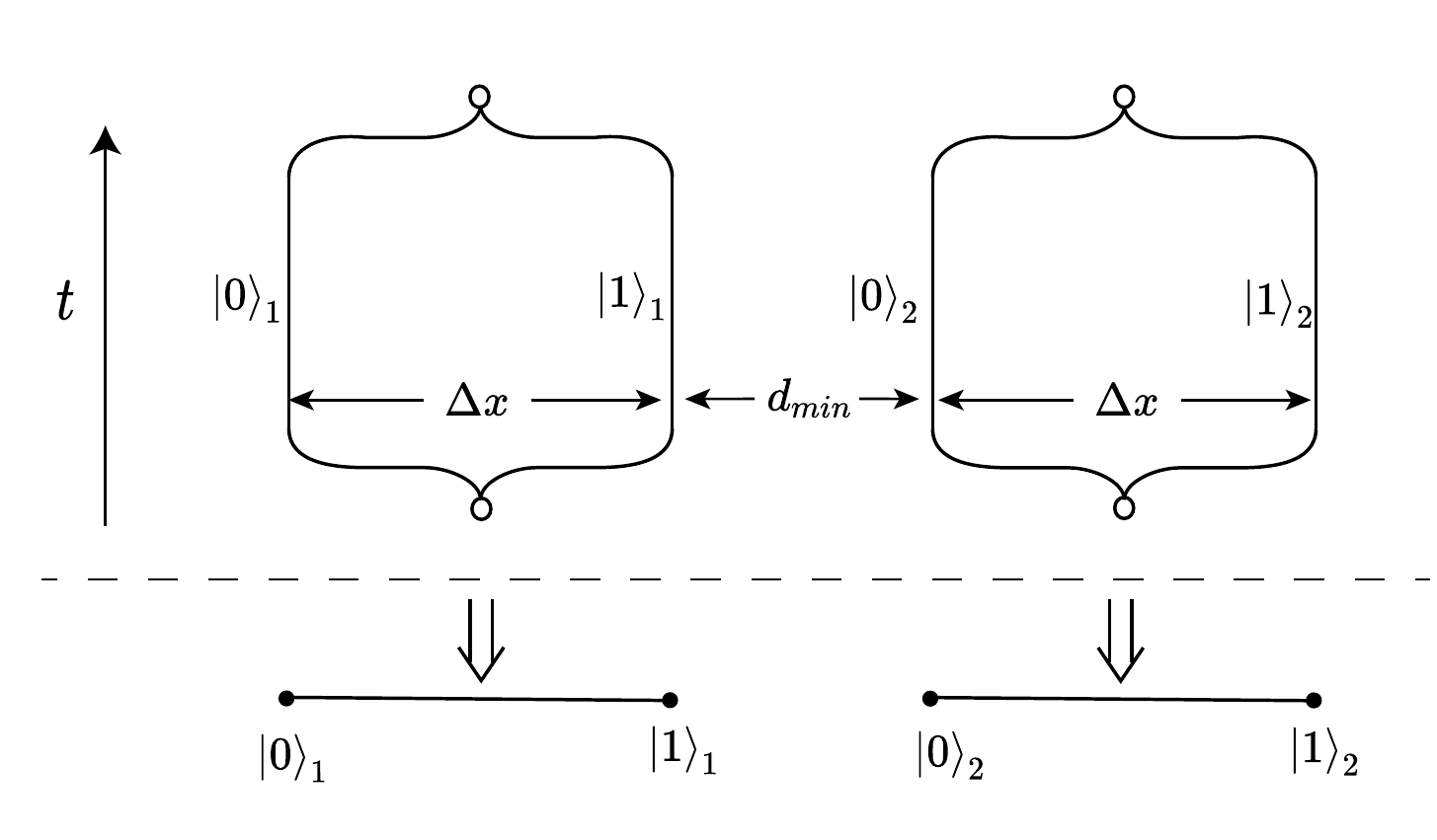}
\caption{Above the dashed line is the plot of original QGEM experiment setup, while below the dashed line is the plot of simplified sketch without time direction: Two particles A and B are placed in the linearly nearby positions, and each particle is prepared in the  superposition of two approximately coordinate eigenstates $\ket{0}$ and $\ket{1}$. As different quantum states of these two particles will induce different gravitational potential, it will generate nontrivial phase differences for states during the evolution of this system and eventually it will generate observable entanglement between two particles when the time of evolution is sufficiently long.}
\label{fig:QGEM}
\end{center}
\end{figure}

In the original QGEM setup, two SG devices are placed in linearly adjacent positions, as illustrated in Fig. (\ref{fig:QGEM}). Later in  \cite{Nguyen:2019huk} a configuration is considered with two SG devices which are placed in a parallel way, and it shows that the parallel setup is more efficient to generate entanglement than the linear one. The setup of two SG devices with different angles is investigated sequently in \cite{Tilly:2021qef}. Furthermore, various configurations in QGEM setup  with three massive particles are considered in \cite{Schut:2021svd}. It is shown that three-qubit system usually generates entanglement more efficient than two-qubit system and behaves better in resisting decoherence. In particular, among all three-qubit cases the configuration with SG devices placed in parallel is still the most efficient setup. Therefore, the three-qubit setup with SG devices in parallel  is the most efficient setup to generate entanglement by now \footnote{The amount of entanglement is measured by the entanglement entropy and compared for all the devices with the same experiment parameters, namely the same distance between two splits of matter wave for each particle, and the same minimal distance between any two matter waves associated with particles. Moreover, the minimal distance is required to avoid the Casimir force \cite{Casimir:1948dh}.}. Inspired by the above work, we intend to consider whether one could continually improve the entanglement generation rate by adding the number of particles $n$, and specially whether the configuration with SG devices in parallel remains to be the best setup with the most efficient generation rate of entanglement, as in cases of n=2 and n=3.
In this paper we systemically analyze the case of n=4 and  find that the setup with SG devices in the form of triangular prism with a center is the most efficient one among all the possible configurations with $n=4$. Furthermore, we argue that this result can be extended to a system with multiple particles up to seven, where the entanglement efficiency is also enhanced in comparison with the setup with fewer particles .

We arrange our paper as following. In next section, we will briefly review QGEM experiment proposal with three-qubit particles, focusing on the improvement of entanglement generation. In section three, we will consider the system with four-qubit particles in detail and show that the configuration in the form of triangular prism with a central particle is the most efficient setup to generate entanglement. We will also extend this setup up to $n=7$ and compare the generation rate of entanglement for different setups. Our summary and conclusions are given in the last section, with some remarks on its novelty and plausibility. Some details on the computation of entanglement are given in the Appendix.

\section{The setup with three massive particles}
In this section we will firstly present the logic line for the computation of entanglement entropy between one specified particle with the other particles in a system composed of $n$ particles, which is the base for witnessing gravity-induced entanglement in QGEM experiment. We will also briefly review the setup with three massive particles and consider all the possible configurations when the parameters are fixed, among of which three configurations have previously been considered in Ref.(\cite{Schut:2021svd}).

\begin{figure}
\begin{center}
\includegraphics[width=0.9\textwidth]{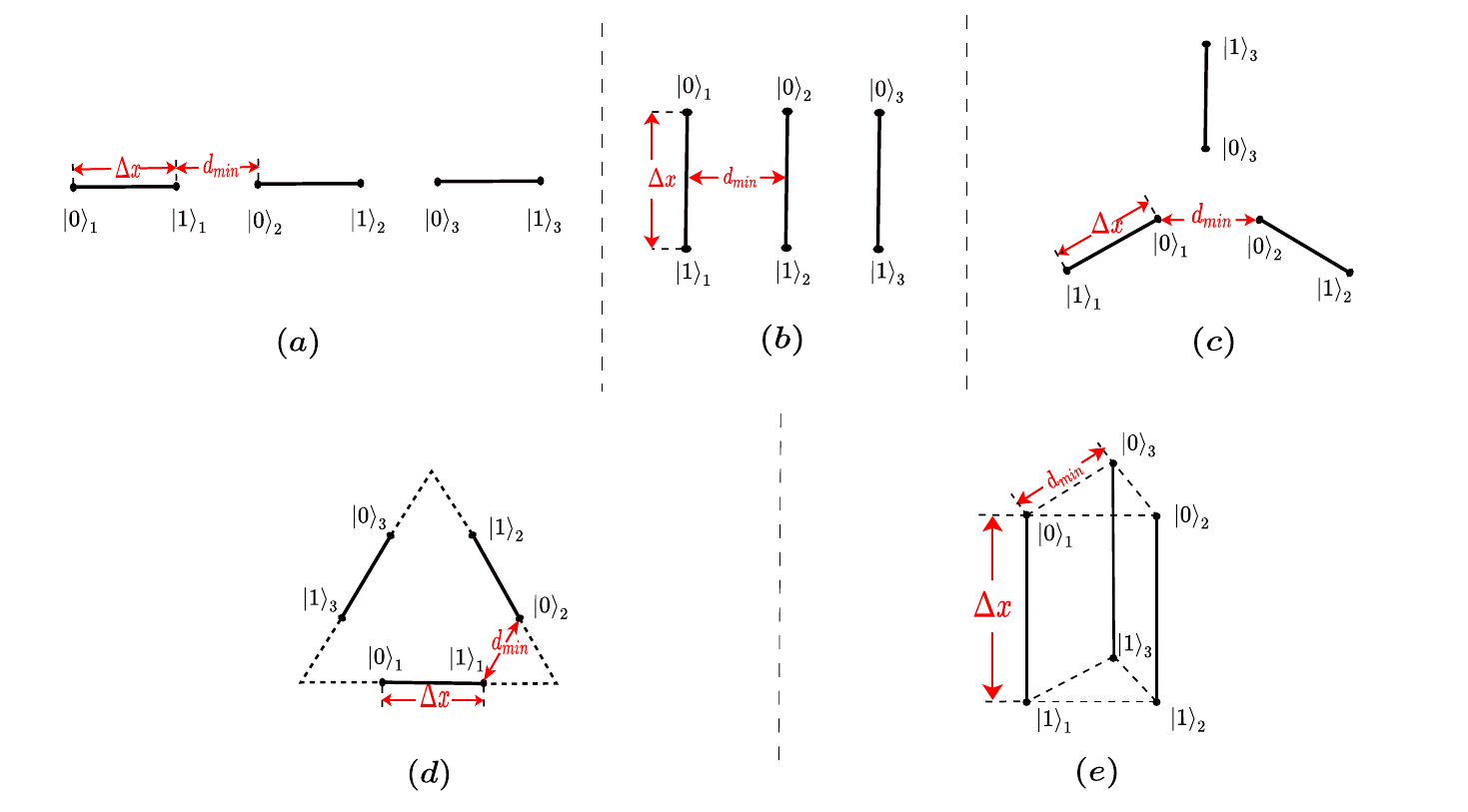}
\caption{Different QGEM setups with three massive particles. 
Each solid line represents a particle, which is in a superposition of $\ket{0}$ and $\ket{1}$. The $i$th particle is in a superposition of $\ket{0}_i$ and $\ket{1}_1$. $\Delta x$ as well as $d_{min}$ has the same value in all setups.
}
\label{fig:n=3setup}
\end{center}
\end{figure}

Let us consider a system composed of $n$ massive particles interacted by gravity due to Newtonian potential, each of which can be viewed as a qubit with two eigenstates of spin, namely $\ket{0}$ and $\ket{1}$. Now we set the initial state of the system as 
\begin{equation}
    \ket{\psi(t=0)} = \bigotimes_{i=1}^n \frac{1}{\sqrt{2}}\qty(\ket{0_i} + \ket{1_i}).
\end{equation}

Next we consider the evolution of the system under the Hamiltonian which is given by 
\begin{equation}
    \hat{H} = \sum_{1 \le i < j \le n} \hat{V}_{ij},
\end{equation}
where $\hat{V}_{ij}$ is the gravitational potential between the $i$th particle and the $j$th particle, and
\begin{equation}
    \hat{V}_{ij} = -G m^2 \mqty(\dmat[0]{1/R({\ket{0_i}, \ket{0_j}}), 1/R({\ket{0_i}, \ket{1_j}}), 1/R({\ket{1_i}, \ket{0_j}}), 1/R({\ket{1_i}, \ket{1_j}})}),
\end{equation}
where $R(\ket{a_i}, \ket{b_j})$ represents the distance between the $i$th particle in the $\ket{a_i}$ state and the $j$th particle in the $\ket{b_j}$ state, and $a_i, b_j = 0$ or $1$.

Then, under the action of  the time evolution operator, it is straightforward to obtain the state at time t as:
\begin{equation}
    \ket{\psi(t)} = e^{-\frac{i}{\hbar}\hat{H}t} \ket{\psi(0)} = \qty(\frac{1}{\sqrt{2}})^n\sum_{i_1, \ldots ,i_n=0,1} e^{-\frac{i}{\hbar} \phi_{i_1 \ldots i_n} t} \ket{i_1 \ldots i_n},
\end{equation}
where the phase $\phi$ is determined by the Newtonian potential as 
\begin{equation}
    \phi_{i_1 \ldots i_n} = - \sum_{1 \le j < k \le n} \frac{G m^2}{R(\ket{i_j}, \ket{i_k})}.
\end{equation}
For various configurations under consideration, we present the specific expressions for  $R(\ket{i_j}, \ket{i_k})$ in Appendix \ref{appendix:A}.

Given the quantum state, the entanglement entropy between particle $i$ and the other particles is given by
\begin{equation}
    S_i = S(\rho_i) = - \Tr\qty(\rho_i \ln \rho_i) = -\sum_j \lambda_j \ln \lambda_j
\end{equation}
where $\rho_i = \Tr_{1, \cdots, \hat{i}, \cdots, n}(\rho)$ is the reduced density matrix of the particle $i$, and $\lambda_j$ are the eigenvalues of $\rho_i$. 

\begin{figure}
\begin{center}
\includegraphics[width=\textwidth]{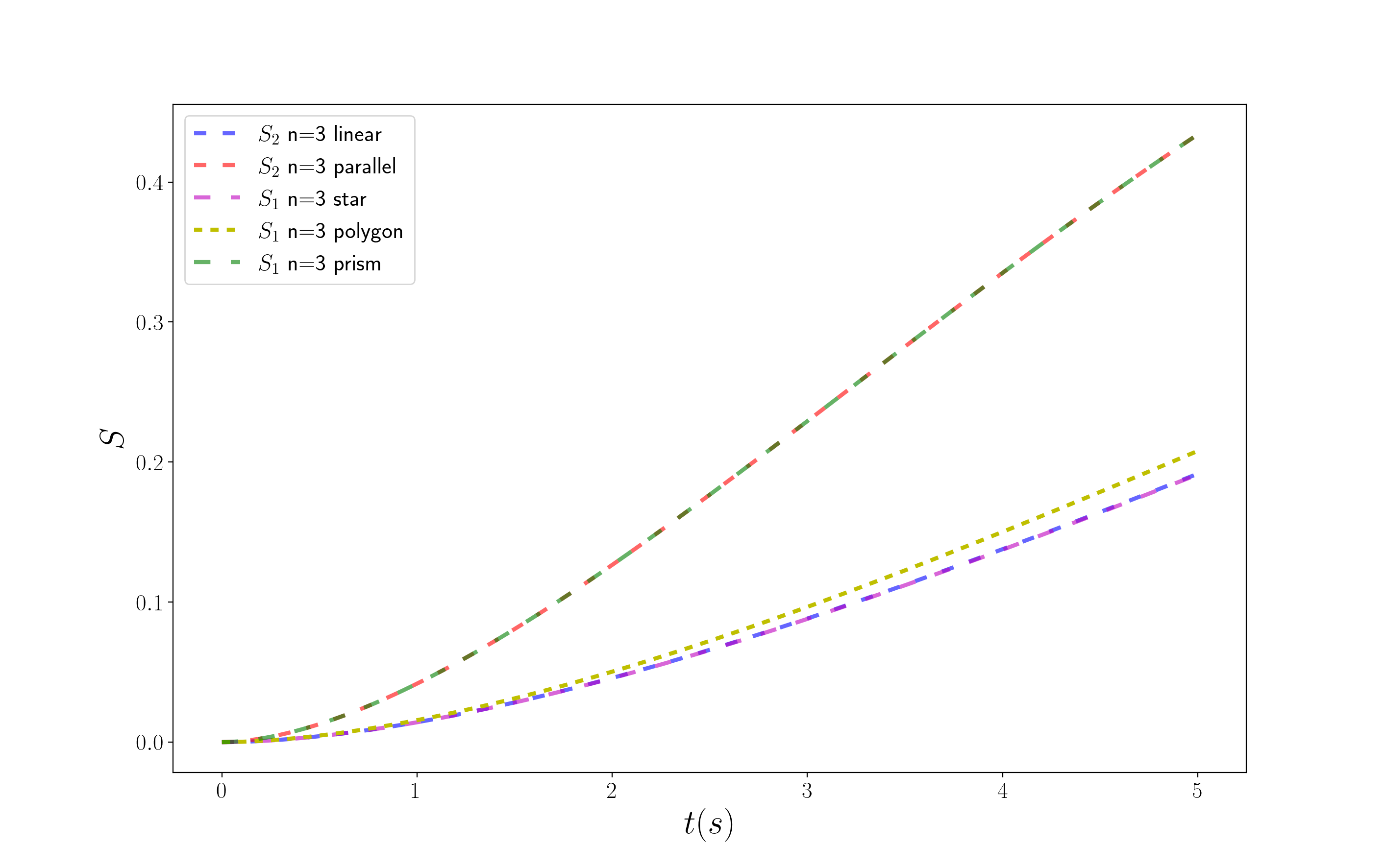}
\caption{The evolution of entanglement entropy in different setups with three massive particles.  Notice that the line of $S_2$ in parallel setup overlaps with the line of $S_1$ in prism setup. And the line of $S_2$ in linear setup almost overlaps with the line of $S_1$ in star setup.}
\label{fig:n=3EE}
\end{center}
\end{figure}

Next we consider the setup with three massive particles and their possible configurations.  First of all, we fix the parameters for all the configurations within the feasible range as discussed in \cite{Bose:2017nin}. We choose the mass of each particle to be $m=10^{-14}kg$. We keep the distance between two channels of each SG device (which correspond to state $\ket{0}_i$ and $\ket{1}_i$ respectively) to be  $\Delta x=250\mu m$, and the minimal distance between any two adjacent SG devices to be $d_{min}=200\mu m$. That is to say, $m$, $\Delta x$ and $d_{min}$ are the same in numerical analysis for all configurations. For $n=3$, it turns out that there are five typical configurations, as illustrated in Fig. (\ref{fig:n=3setup}), among of which three configurations (a), (b) and (c) have previously been considered in Ref.(\cite{Schut:2021svd}), which may be called as  the linear, the parallel and the star setup, respectively. It is pointed out  in Ref.(\cite{Schut:2021svd}) that the second particle in the parallel setup has the highest rate of entanglement generation, which reaches the observable criteria within 5 seconds, followed by the star setup, and then the linear setup. As a matter of fact we may construct other two configurations, as illustrated in (d) and (e) \footnote{We ignore the impact of the gravity due to the Earth on all the particles.}. We call them  the polygon setup and the prism setup, respectively. In Fig. (\ref{fig:n=3EE}), we demonstrate the highest generation rate of the entanglement for all the five configurations. For linear and parallel setup, the particle in the middle has the highest rate of entanglement generation. This is because this  particle has a smaller average distance to the other particles and therefore has a stronger interaction with the other particles. For the star, the polygon and the prism setup, because of the symmetry, we just need to calculate the entropy for $S_1$. From this figure, we notice that the linear and star setup have the slowest generation rate of entanglement, while the polygon setup is slightly faster, and both the parallel and prism setup have the highest rate. This is not surprising since  the middle particle in parallel setup experiences the same Newtonian potential as any particle in prism setup.

As a summary, in both cases of $n=2$ and $n=3$, the parallel setup exhibits the largest rate of entanglement generation in comparison with the other setups with the same number of particles. We are wondering if this would be true when the number of particles increases, and if the entanglement efficiency is enhanced in comparison with the setup with fewer particles. We will investigate these issues in next section. 

\section{The setup with four and more massive particles}
In this section we firstly consider the generation rate of entanglement for four massive particles in detail, and then generalize it to cases with more massive particles. 

\begin{figure}
\begin{center}
\includegraphics[scale=0.9]{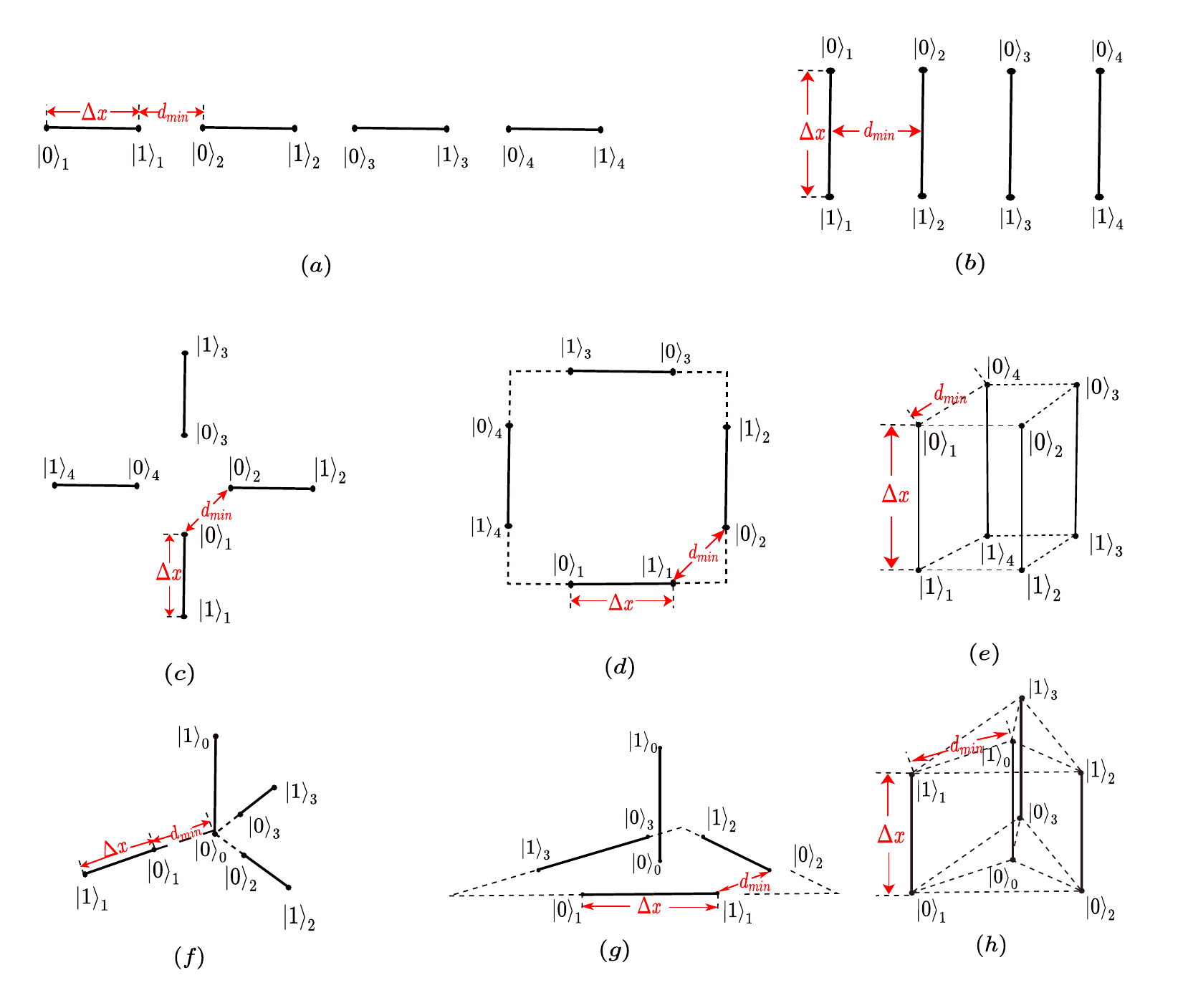}
\caption{Different QGEM setups with four massive particles. The setups (a)-(e) are direct extension of Fig.\ref{fig:n=3setup}(a)-(e), while setups (f), (g), (h) are obtained by adding a central particle to Fig.\ref{fig:n=3setup}(c), (d), (e).}
\label{fig:n=4setup}
\end{center}
\end{figure}

First of all, for a system with four massive particles, all the configurations we considered in previous section can be extended, as illustrated in Fig. \ref{fig:n=4setup} (a)-(e). Furthermore, for the star, polygon and prism setups, we may also obtain a new configuration by inputting the fourth particle at the center of the setup respectively, as illustrated in Fig.\ref{fig:n=4setup} (f), (g), (h), where the SG device of the fourth particle is perpendicular to the plane formed by  SG devices of the other three particles in (f) and (g), while  in (h) the SG device of the fourth particle is parallel to the prism. Totally, one obtains eight possible configurations for the QGEM setup for four massive particles.

\begin{figure}
\begin{center}
\includegraphics[width=\textwidth]{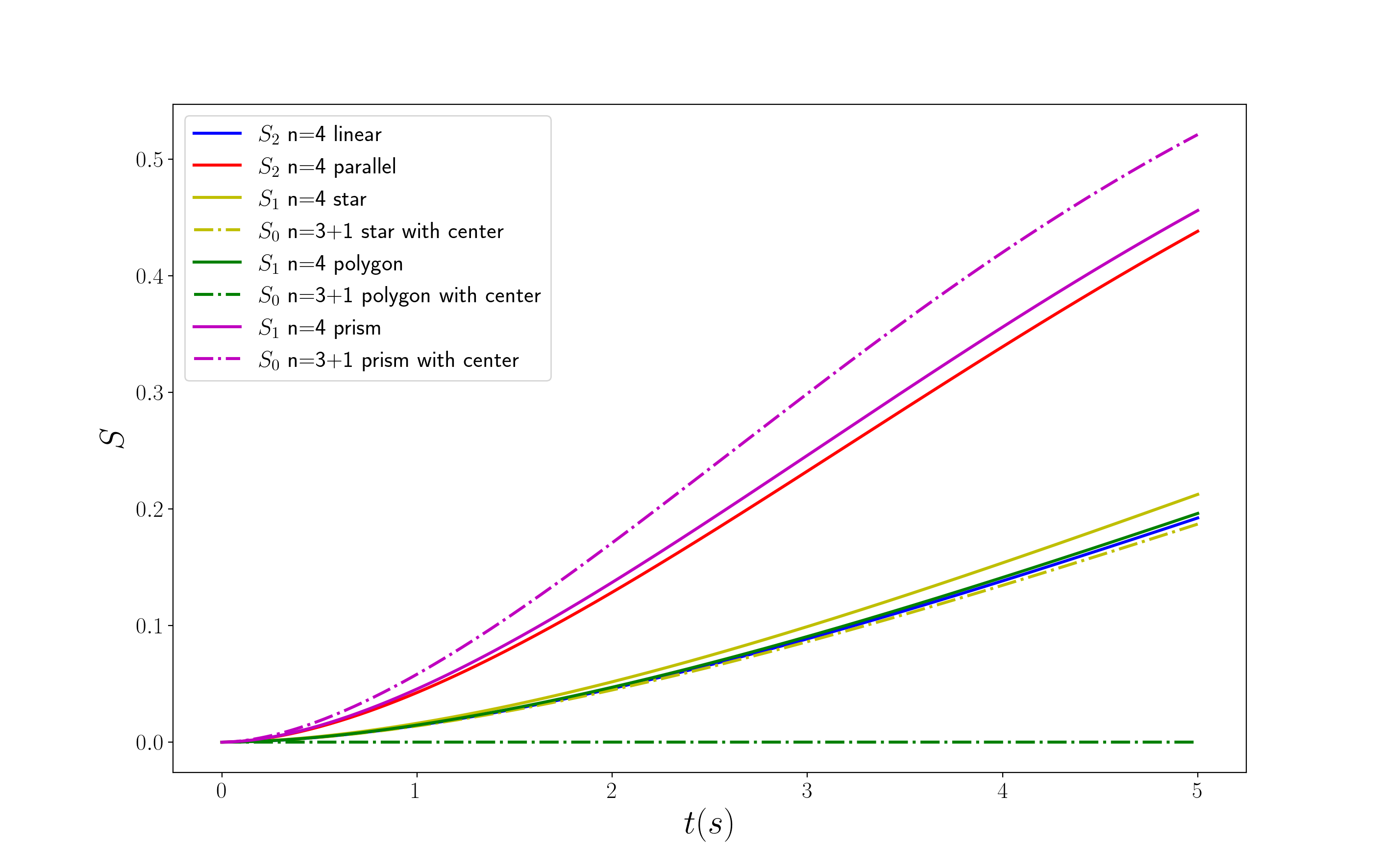}
\caption{The evolution of the entanglement entropy in different setups when $n=4$. }
\label{fig:n=4EE}
\end{center}
\end{figure}

Now we are concerned with the generation rate of the entanglement in these different configurations. Again, we only need to consider $S_2$ for the linear and parallel setup since it exhibits the most efficient rate among all the particles, and consider $S_1$ for the star, polygon and prism setup due to the symmetry of the configuration. For the setups with a central particle, namely (f), (g) and (h) in Fig. (\ref{fig:n=4setup}),  we only need to consider the central particle which has the minimal average distance and thus exhibits the highest entanglement generation rate among all the massive particles. We denote the entanglement entropy of the central particle as $S_0$. We perform the numerical analysis and illustrate the results for the highest generation rate in each configuration in Fig. (\ref{fig:n=4EE}).

Most importantly, from Fig.\ref{fig:n=4EE}  it is clear to see that the central particle in the setup of triangular prism $(h)$ exhibits the largest generation rate of entanglement among all the configurations, since it has the minimal distance to all the other three particles. Secondly, the prism setup is better than the parallel setup when $n=4$, although both setups have the same entanglement generation rate in the case of $n=3$. In addition, for the polygon setup with central particle $(g)$, $S_0$ is always zero due to the symmetry of the setup. We present the proof of this statement in Appendix \ref{appendix:B}.

The above numerical results are consistent with what we expect. As we mentioned above, in the parallel setup of $n=4$, the fourth particle is farther away from the second particle, so the gravitational interaction between the fourth particle and the second one becomes weak, contributing little to the entanglement entropy. On the other hand, in the prism setup, the average distance between particles is smaller than that in the parallel setup, thus the generation rate of entanglement is  improved more efficiently when adding particles. In particular, if we add a particle at the center of the prism, the average distance between this particle and all the other particles is the smallest, thus exhibiting the largest generation rate of entanglement entropy.

\begin{figure}
\begin{center}
\includegraphics[scale=1]{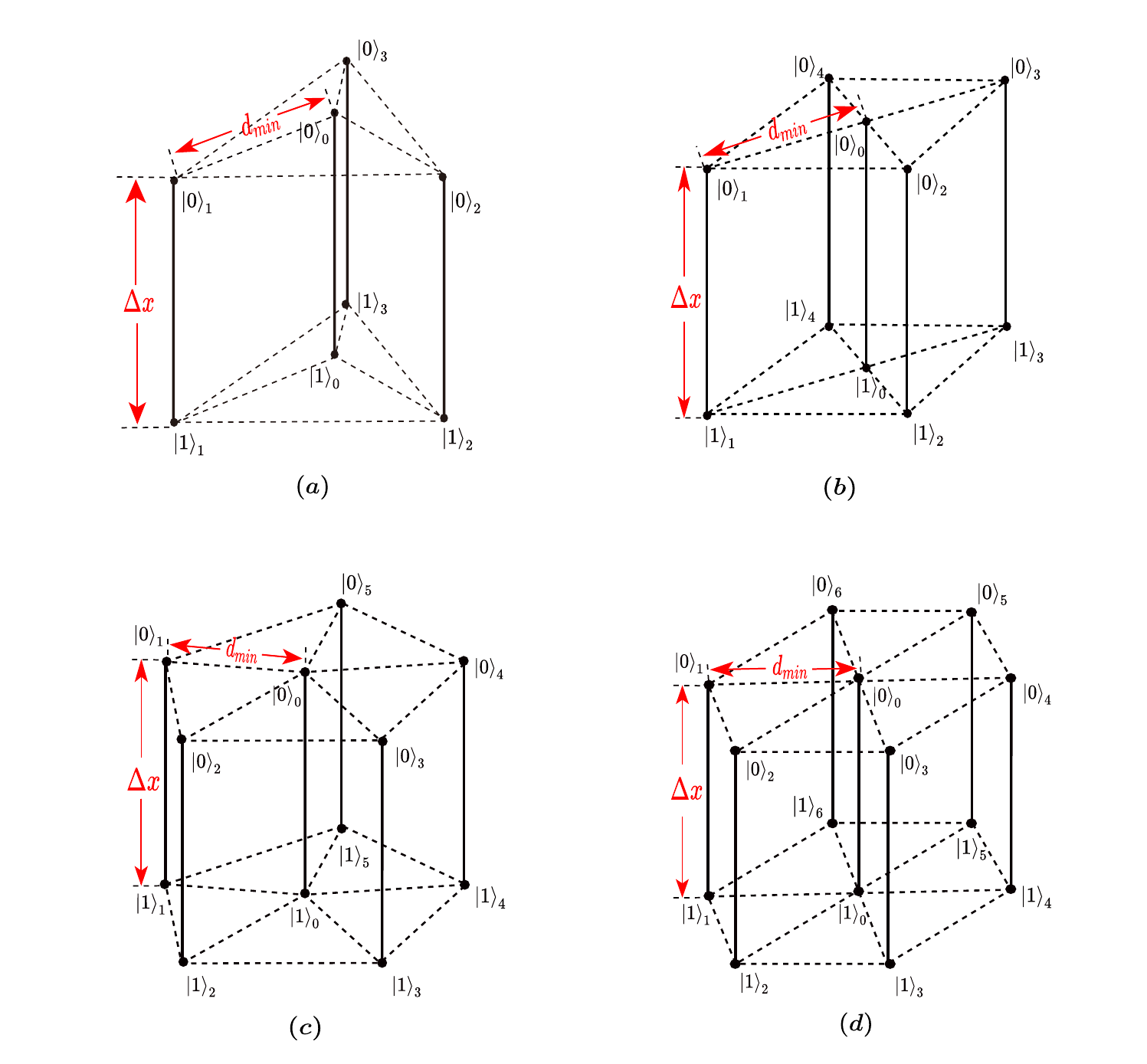}
\caption{The prism setup with a central particle for $n=4, 5, 6, 7$.}
\label{fig:n=4567setup}
\end{center}
\end{figure}

\begin{figure}
\begin{center}
\includegraphics[width=\textwidth]{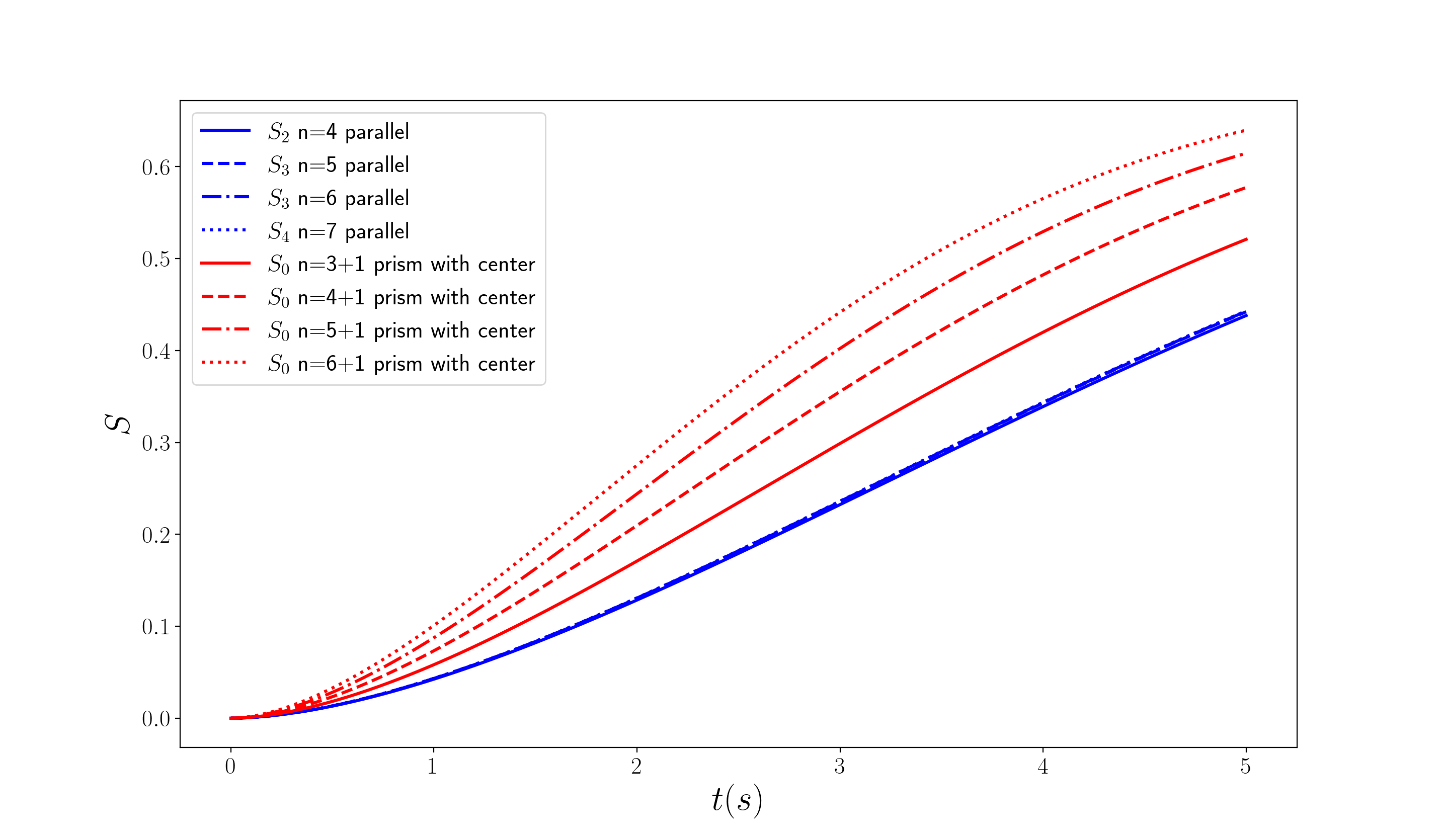}
\caption{The time evolution of entanglement entropy in parallel setup and prism setup with a central particle when $n=4, 5, 6, 7$. Note that the generation rate of entanglement in prism setups increases with the number of particles dramatically, while all the four curves representing the evolution of entanglement entropy in parallel setups almost overlap.}
\label{fig:n=4567EE}
\end{center}
\end{figure}

Inspired by the above analysis on $n=4$, next we intend to generalize it by considering setups with more particles and expect that the prism setup with a central particle will exhibit the maximal generation rate of entanglement. We depict the prism configuration for $n=4, 5, 6, 7$ in Fig. (\ref{fig:n=4567setup}). We find that these configurations with a central particle always exhibit the highest generation rate of entanglement among all the possible configurations with the same number of particles. More importantly, we find that this rate increases dramatically as the number of particles increases, as illustrated in Fig. (\ref{fig:n=4567EE}). In contrast, one finds that for other configurations the generation rate is barely improved when the number of particles increases. For instance we show the rate of the parallel setup with different number of particles in Fig.(\ref{fig:n=4567EE}) as well. The rate is almost saturated since the increased particles are far from the observed particle such that their effects may be negligible.  

Based on the above investigation, we conclude that for $n>3$ the prism setup with a central particle exhibits the maximal generation rate of entanglement and this rate can be improved by increasing the number of particles up to $n=7$. When the number of particles is larger than seven, then the distance between the central particle and other particles would not be the minimal one, but larger than the distance between any two neighboring particles surrounding the central particle. We will not consider these cases from the actual situation of the experiment.

\section{Conclusion and Discussion}

With the goal of constructing the setup with the most efficient rate of entanglement generation due to the gravity interaction, we have investigated new QGEM setups with the number of massive particles $n>3$. For $n=4$, we have found that the particle at the center of the prism setup exhibits the most efficient entanglement behavior since it has the most pairs of neighboring particles. This situation is in contrast to the cases of  $n=2$ and $n=3$ which have previously been studied in literature. For $n=2$ and $n=3$, the parallel setup exhibits the best entanglement behavior in comparison with other setups.Furthermore, we have extended  this strategy to the setups with more particles up to $n=7$, where the prism setup with center is always the best one for entanglement generation since the distance between the central particle with other particles remains the minimal among all the distances between any two particles. We have found that the efficiency for entanglement generation can also be considerably improved with the increase of the number of particles, in contrast with what happens in the parallel setup, where the generation rate becomes saturated after $n>4$. This work has provided the strategy to construct the QGEM setup with the best generation rate of entanglement. To guarantee that these new proposed setups may be implemented in experiments, we give some remarks on the relevant properties of these setups.

\begin{enumerate}
\item{The optimization of parameters.}
As an improvement of the original QGEM setup, the prism setup with center proposed in this paper could generate entanglement more rapidly so it may be helpful to loose the constraints for the mass of the particles as well as the time of the coherence-preserving, which are the key parameters for real lab experiment. For instance, when $t=5s$, the entanglement entropy of the central particle of n=3 in parallel setup (which exhibits the most efficient generation rate of entanglement in $n=3$) is about 0.43, while the entanglement entropy of the central particle in $n=4$ prism setup with center may reach the same value just in $t=4s$. Moreover, the entanglement entropy of the central particle in $n=7$ prism setup with center reaches 0.43 just in about $t=2.8s$, which is approximately an half of $5s$. Alternatively, if the time of the coherence-preserving remains to be five second, then one could loose the requirement of the mass of the matter wave from $10^{-14}kg$ to a smaller quantity, which would make the experiment easier.

\item{Decoherence.} Decoherence is an important issue that should be taken into account seriously in such QGEM experiments. While following the analysis on decoherence in \cite{Schut:2021svd}, it is more likely that the setup with more particles is better resilient to the decoherence.
In hence, the decoherence effects should not be an obstacle to prevent us from adding more particles in QGEM setup.

\item {Entanglement witness.} In prism setup with center, with the increase of the generation efficiency of entanglement, one has to pay price to have more time consumption for measuring quantum states. That is to say, the setup with more massive particles would cost more time to measure the quantum states to compute the entanglement in experiment \cite{Schut:2021svd}. Anyway, this may not be an unconquerable obstacle, and we expect it might be improved by some new entanglement witness prescription as discussed in \cite{Guff:2021mfw}. 

\end{enumerate}

\section*{Acknowledgments}

We are very grateful to Kai Li, Yuxuan Liu and Menghe Wu for helpful discussions on QGEM experiment. We acknowledge the support from the Innovative Projects of Science and Technology at IHEP. 
This work is supported in part by the Natural Science Foundation
of China under Grant No.~11875053, 12035016 and 12275275. It is also supported by Beijing Natural Science Foundation under Grant No. 1222031.

\appendix
\section{Distance of different setups}
\label{appendix:A}
\renewcommand{\d}{d_{min}}
\newcommand{\x}{\Delta x}
\renewcommand{\r}[1]{R^{(\text{#1})}(\ket{i_j}, \ket{i_k}; \d, \x)}
\newcommand{\rcenter}[1]{R^{(\text{#1 with center})}(\ket{i_0}, \ket{i_k}; \d, \x)}

In this appendix we present the expressions for the distance between two massive particles in various setups. For linear setup, if the $j$th particle lies in $\ket{i_j}$ state, and the $k$ particle is in $\ket{i_k}$ state, then the distance between these two particles is denoted as $R(\ket{i_j}, \ket{i_k}; \d, \x)$, which is given by
\begin{equation}
    \r{linear} = (k-j)(\d + \x) + (i_k - i_j) \x
,\end{equation}
where we assume that $j < k$, without loss of generality.

For parallel setup, $R$ is 
\begin{equation}
    \r{parallel} = \sqrt{\qty[(k-j)\d]^2 + \qty[(i_k - i_j)\x]^2}.
\end{equation}

For star setup, $R$ is
\begin{gather}
    \r{star} = \sqrt{a^2 + b^2 - 2 a b \cos((k-j)\frac{2\pi}{n})},
\end{gather}
where $a = \frac{\d}{2 \sin(\frac{\pi}{n})} + i_j \x$ and $b = \frac{\d}{2 \sin(\frac{\pi}{n})} + i_k \x$.

For star setup with a central particle, the central particle is regarded as the $0$th particle, and $R$ is given by
\begin{gather}
    \rcenter{star} = \sqrt{(\d+i_k \x)^2 + (i_0 \x)^2},\\
    \r{star with center} = R^{(\text{star})}(\ket{i_j}, \ket{i_k}; 2\d \sin\frac{\pi}{n}, \x),
\end{gather}
where $j, k \ge 1$.

For polygon setup, $R$ is 
\begin{gather}
    \r{polygon} = l \sqrt{2(1 - \cos\qty[\frac{2\pi}{n}(k-j) + 2(i_k - i_j)\theta])},
\end{gather}
where $\theta = \arcsin \frac{\x}{2l}$ and $l = \sqrt{\frac{\d^2 + 2\d\x\cos(\frac{\pi}{n}) + \x^2}{4 \sin^2(\frac{\pi}{n})}}$. 

And for polygon setup with a central particle, $R$ is 
\begin{gather}
    \rcenter{polygon} = l,\\
    \r{polygon with center} = R^{(\text{polygon})}(\ket{i_j}, \ket{i_k};\d, \x).
\end{gather}

For prism setup, $R$ is
\begin{equation}
    \r{prism} = \sqrt{\qty[\d \frac{\sin((k-j)\frac{\pi}{n})}{\sin \frac{\pi}{n}}]^2 + \qty[(i_j - i_k)\x]^2},
\end{equation}

And for prism setup with a central particle, $R$ is
\begin{gather}
    \rcenter{prism} = \sqrt{d^2 + \qty[(i_0 - i_k)\x]^2},\\
    \r{prism with center} = R^{(\text{prism})}(\ket{i_j}, \ket{i_k}; 2\d \sin\frac{\pi}{n}, \x).
\end{gather}

\section{The proof of $S_0 = 0$ for polygon setup with a central particle}
\label{appendix:B}
The state is given by
\begin{align*}
    \ket{\psi(t)} &= N\sum_{i_0\ldots i_n}e^{-\frac{i}{\hbar}\phi_{i_0\ldots i_n} t}\ket{i_0\ldots i_n}\\
    &= N\sum_{i_1\ldots i_n}\qty(e^{-\frac{i}{\hbar}\phi_{0,i_1\ldots i_n} t}\ket{0,i_1\ldots i_n}+e^{-\frac{i}{\hbar}\phi_{1,i_1\ldots i_n} t}\ket{1,i_1\ldots i_n}),
\end{align*}
where $N$ is a normalization factor, and $N = \qty(\frac{1}{\sqrt{2}})^n$. Note that $R(\ket{0_0},\ket{i_k})$ is the same for all $\ket{i_k}$, thus we have
\begin{align*}
    \phi_{0,i_1\ldots i_n} &= -G m^2 \qty(\sum_{k=1}^n \frac{1}{R(\ket{0_0}, \ket{i_k})} + \sum_{1 \le j < k \le n}\frac{1}{R(\ket{i_j}, \ket{i_k})})\\
    &= -G m^2 \qty(\frac{n}{\d} + \sum_{1 \le j < k \le n}\frac{1}{R(\ket{i_j}, \ket{i_k})})\\
    &= C_1 + \phi_{i_1 \ldots i_n}.
\end{align*}
And also $\phi_{1,i_1\ldots i_n} = C_2 + \phi_{i_1 \ldots i_n}$. Then
\begin{align*}
    \ket{\psi(t)} &= N \qty(e^{-\frac{i}{\hbar} C_1 t} \ket{0} + e^{-\frac{i}{\hbar} C_1 t} \ket{1}) \otimes \qty(\sum_{i_1 \ldots i_n} e^{-\frac{i}{\hbar} \phi_{i_1 \ldots i_n} t} \ket{i_1 \ldots i_n}) \\
    &= \ket{\phi(t)}\otimes\qty(\sum_{i_1 \ldots i_n} e^{-\frac{i}{\hbar} \phi_{i_1 \ldots i_n} t} \ket{i_1 \ldots i_n}).
\end{align*}
Therefore, $\rho_0 = \Tr_{1\ldots n} (\ket{\psi(t)}\bra{\psi(t)}) = \ket{\phi(t)}\bra{\phi(t)}$ is a pure state, and $S(\rho_0) = 0$.

\end{document}